# Dielectric response in twisted $MoS_2$ bilayer facilitated by spin-orbit coupling effect

Yu-Hao Shen[1,2], Jun-Ding Zheng[3,5], Wen-Yi Tong[3,5], Zhi-Qiang Bao[3,5], Xian-Gang Wan[1,2*], Chun-Gang Duan[3,4,5*]

[1]National Laboratory of Solid-State Microstructures and School of Physics, Nanjing University, Nanjing 210093, China

[2]Collaborative Innovation Center of Advanced Microstructures, Nanjing University, Nanjing 210093, China

[3]Key Laboratory of Polar Materials and Devices, Ministry of Education, East China Normal University, Shanghai 200241, China

[4]Collaborative Innovation Center of Extreme Optics, Shanxi University, Taiyuan, Shanxi 030006, China.

[5]Shanghai Center of Brain-inspired Intelligent Materials and Devices, East China Normal University, Shanghai 200241, China

*Author to whom any correspondence should be addressed.

E-mail:

cgduan@clpm.ecnu.edu.cn

xgwan@nju.edu.cn

## ABSTRACT

Twisted van der Waals bilayers offer ideal two-dimensional (2D) platforms for exploring the intricate interplay between the spin and charge degrees of freedom of electrons. By investigating twisted $MoS_2$ bilayer, featuring two distinct stackings but with identical commensurate supercell sizes, we reveal an unusual dielectric response behavior inherent to this system. Our first-principles calculations demonstrate that the application of an out-of-plane electric field gives different responses in electronic polarization. Upon further analysis, it becomes clear that this dielectric response comes from the planar charge redistribution associated with spin-orbit coupling (SOC) effect. The underlying mechanism lies in the fact that the external electric field tends to modify the internal pseudo-spin texture $\boldsymbol{\sigma}$, subsequently generating an out-of-plane (pseudo-) spin current $\boldsymbol{j}_s \propto \boldsymbol{\sigma} \times \boldsymbol{B}_R$ as response to an in-plane Rashba SOC field $\boldsymbol{B}_R$. It is found that the generated $\boldsymbol{j}_s$ is opposite for the two distinct stackings, resulting in opposite in-plane electric susceptibility. As a consequence, through magnetoelectric coupling within such nonmagnetic system, there give rise to opposite tendency to redistribute charge, ultimately leading to an amplified or suppressed dielectric response.

## INTRODUCTION

Twisted by an angle with respect to each other, two van der Waals separated layers stack superlattice with moiré period[1-7], which brings about rich physics in the study of low energy quasi-particle excitation of electron[8-11]. Under patterned modulation of both intralayer and interlayer potential of a typical bilayer system, there exhibits exotic and unique electric and magnetic properties[5,11-18]. Two individual monolayers stacked in twisted bilayer structures are compatible with moiré potential[19-22], which is associated with the local symmetry of the two relatively sliding and rotated layers [5,23-26]. When an out-of-plane electric field is applied, it induces a planar redistribution of charge in the bilayer system[17,26], since the atomic sites of the two layers are not aligned with each other. In absent of mirror symmetry of *xy*-plane, there shifts the charge center of such bilayer and creates an effective electric dipole, resulting in both interlayer and intralayer charge transfer. However, when subject to atomic SOC effect, the formation of electric dipoles may rely on the interplay between the extrinsic pseudo-spin pattern (Rashba SOC effect) by an out-of-plane electric field and the intrinsic one, as reported in our previous work[17].

In moiré system[16,17,27], the intrinsic pseudo-spin pattern can be attributed to the superlattice structure with reduced symmetry. And we even find a negative dielectric response in a magnetic twisted bilayer, where the creation of electric dipoles relies on both the pseudo-spin texture in momentum space and the magnetic exchange coupling in real space[17]. This is exactly the dielectric response in a magnetic way. For non-magnetic system, however, the spin polarization is purely from SOC, which is so small for graphene system. And in transition metal dichalcogenides (TMDCs) of the 2H-$MX_2$ type (where M=Mo, W and X=S, Se, Te), with their large atomic SOC. Besides, there emerges both band topology and electron-electron correlation when interlayer twist is introduced[19,22,28]. They are ideal platforms for us to explore the dielectric behavior within such nonmagnetic system in analogy to that in magnetic system.

In this paper, we take twisted bilayer $MoS_2$ as an example and perform first-principles calculations included SOC effect. We define a commensurate supercell shown in Fig. 1(a) as R-type like and H-type like stacking. It is shown that by calculating the electronic polarization under an out-of-plane electric field, a notable difference in dielectric susceptibility appears for the two distinct stacking cases. And the calculated results are apparently different from those without SOC included, where the dielectric response is almost no difference. We attribute these observations to the planar charge redistribution with SOC effect considered when applied perpendicular electric field. Further analysis show that taking into account Rashba SOC, an out-of-plane (pseudo-) spin current $\boldsymbol{j}_s \propto \boldsymbol{\sigma} \times \boldsymbol{B}_R$ is generated as response to an in-plane SOC field $\boldsymbol{B}_R$. There gives rise to opposite in-plane electric polarization for the two distinct stacking cases, whereas without SOC the planar dielectric polarization is expected to be nearly identical. It tends to redistribute charge to amplify or resist the external electric field. Our results suggest an intuitive picture how spin and charge polarization of an electron can be entangled.

## COMPUTATIONAL METHOD

### A. First-principles calculations

The first-principles calculations are performed with density functional theory (DFT) using the projector augmented wave (PAW) method implemented in the Vienna ab initio Simulation Package (VASP)[29]. The exchange-correlation potential is treated in Perdew-Burke-Ernzerhof form[30] of the generalized gradient approximation (GGA-PBE) with a kinetic-energy cutoff of 400 eV. A well-converged 7×7×1 Monkhorst–Pack $k$-point mesh is chosen in self-consistent calculations. The convergence criterion for the electronic energy is $10^{-5}$ eV and the structures are relaxed until the Hellmann-Feynman forces on each atom are less than 1 meV/Å. In our calculations, the SOC effect can be included and the dispersion corrected DFT-D2 method[31] is adopted to describe the van der Waals interactions. The external electric field is introduced with the planar dipole layer method and the electronic polarization calculations under applied out-of-plane electric field is based on Berry Phase (BP) method implemented in DFT.

### B. Lattice structure of the two distinct stackings

For the monolayer 2H-$MoS_2$, the intermediate layer of hexagonally arranged Mo atoms are sandwiched between two atomic layers of S. The optimized lattice constant is 3.19 Å. Start from 30° twisted bilayer system, there generates two moiré structure when top layer rotates left or right with respect to bottom one[32]. This kind of chirality defines them as “conjugated” partners.

Two typical commensurate lattices are shown in Fig. 1(a), where the corresponding supercell possesses the $\sqrt{7}\times\sqrt{7}$ times size of the unit cell. There form different interlayer stackings, defined as R-type like stacking and H-type like stacking respectively. The inset figures show the side view of the local high-symmetric structure in the moiré superlattice about, e.g., for H-type like case, Mo(S) atoms of top layer are locally aligned with S(Mo) atoms of bottom layer, as denoted by M(N) stacking sites. And the registry of its “conjugated” structure is just, i.e., the R-type like case, Mo(S) atoms are locally aligned with Mo(S) atoms.

## RESULTS AND DISCUSSION

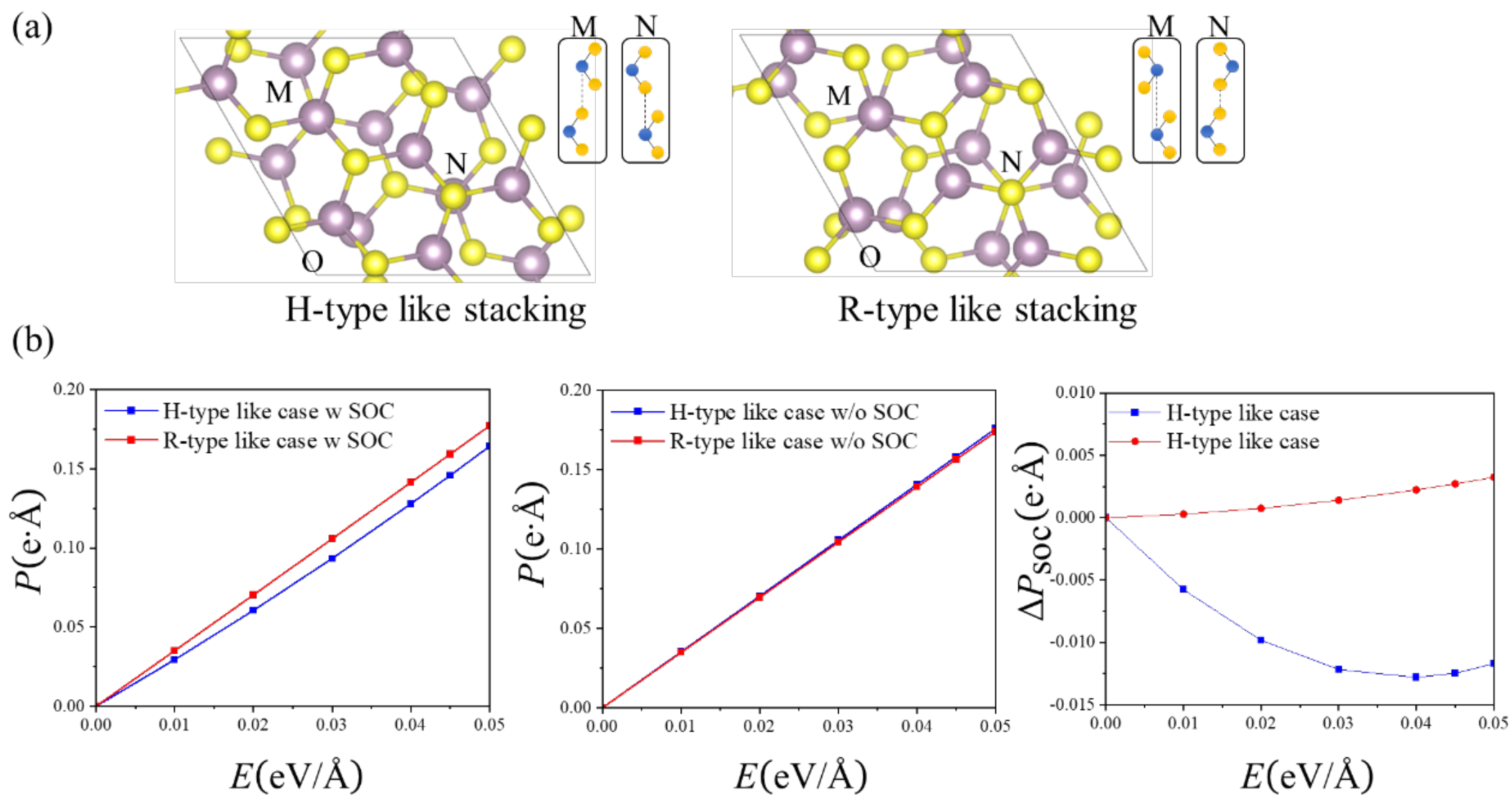

**Figure 1.** (a) Top view of H-type like and R-type like stacking for twisted bilayer $MoS_2$. The purple and yellow balls represent Mo and S atoms, respectively. The inset shows a schematic side view for the local high-symmetric structure in M and N sites of the moiré superlattice. (b) The BP method calculated electronic polarization $P$ and $\Delta P_{\mathrm{SOC}}=P(\text{w SOC})-P(\text{w/o SOC})$ variation as $E$ evolves from 0 to 0.05, in units of eV/Å.

We calculate the electronic polarization under applied out-of-plane electric field $E$ with and without SOC included. Here, the positive field is defined to point from bottom layer to top layer. From the calculated results shown in the left panel of Fig. 1(b), it is evident that as $E$ approaches 0, the susceptibility $\partial P/\partial E$ is apparently suppressed in the H-type case with SOC included, compared to the R-type case. This difference becomes even more apparent in the right panel, which shows that SOC primarily has a minor positive and a significant negative impact on electronic polarization. This is indicated by calculating the polarization difference $\Delta P$=$P$(w SOC)-$P$(w/o SOC).

### A. Calculated pseudo-spin texture

We can understand above exotic dielectric phenomenon by the role of SOC play with electric field. For a local inversion asymmetric structure[33,34], there emerges an effective magnetic field that comes from atomic SOC as $\boldsymbol{B}_{eff} \propto \nabla V \times \boldsymbol{p}$. The associated local electric field $\nabla V$, which depends on the atomic site symmetry, is essential to generate the pseudo-spin pattern in momentum space through the Hamiltonian $H_{pseudo-spin} = \boldsymbol{B}_{eff} \cdot \boldsymbol{\sigma}$. Here, $\boldsymbol{\sigma}$ represents the spin Pauli matrices and we address the spin texture as pseudo-spin texture that corresponds to the specific pseudo-spin Hamiltonian. It differs from a spin Hamiltonian for a magnetic system, where the spin represents real spin. The pseudo-spin defined here indicates the orbital spin polarization of the nonmagnetic system.

As we apply an out-of-plane electric field, there is a tendency for the pseudo-spin pattern to undergo changes due to Rashba SOC. A direct impact on the intrinsic pseudo-spin $\boldsymbol{\sigma}$ under a vortex Rashba SOC field $\boldsymbol{B}_R$ can be a 'spin torque' as $\boldsymbol{\sigma} \times \boldsymbol{B}_R$, resulting in the change of spin density $\Delta\sigma$. It is accompanied by an induced charge density $\Delta\tilde{\rho}$ that purely from SOC effect, besides the electric field induced $\Delta\rho$ directly. This indirect $\Delta\tilde{\rho}$ under applied electric field, in turn, has an influence on the electronic polarization. It motivates us to check the intrinsic pseudo-spin $\boldsymbol{\sigma}$ for the two different stacking cases.

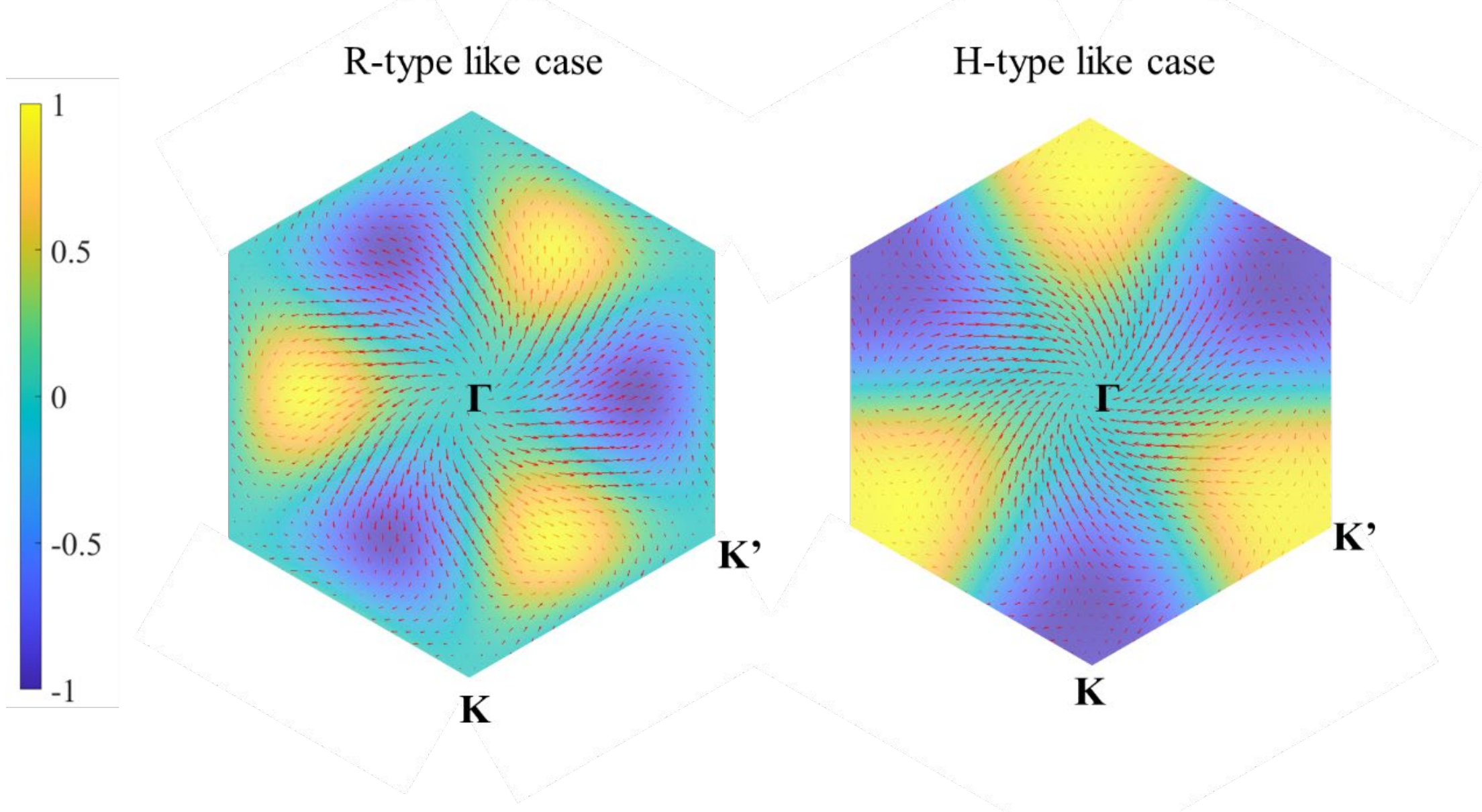

**Figure 2.** Normalized pseudo-spin texture $\langle \boldsymbol{\sigma} \rangle$ corresponds to the two different stacking cases in the absent of applied electric field. The in-plane and out-of-plane components are shown in red arrows and color map, respectively. Here, we sum over all occupied states as we considered a non-vanished spin polarization due to *PT* asymmetry.

Since we need to explore the SOC effect on the electronic polarization, the pseudo-spin texture $\langle \boldsymbol{\sigma} \rangle$ is calculated from sum over all occupied states. Yet, note that for a two-by-two pseudo-spin $H_{pseudo-spin} = \boldsymbol{B}_{eff} \cdot \boldsymbol{\sigma}$, the sum of $\langle \boldsymbol{\sigma} \rangle$ over spin $\boldsymbol{\sigma}$ space vanishes due to the *PT* symmetry, where a Zeeman-type spin splitting term $B_{eff,z}\sigma_z$ is not allowed. Here, *P* denotes the inversion symmetry (*P*) and *T* is time-reversal symmetry. For a nonmagnetic 2D system, if *P*-symmetry is absent (e.g., an in-plane dipole is included), there usually exists a Zeeman-type polarization of the components $\langle \sigma_z \rangle$ [34]. And if we sum over all the states in spin space, there results in non-vanished spin polarization. In our cases, for both two superlattice, they exhibit $D_3$ point group symmetry with respect to the rotation center, which is generated by a three-fold rotation $C_{3z}$ and a two-fold rotation $C_{2y}$/$C_{2x}$ so that the *P*-symmetry is absent. As shown in the Fig. 2, the maximum (minimum) of $\langle \sigma_z \rangle$ occurs in **Γ**-**M** and **K** respectively for H-type and R-type cases. However, there leads to inward and outward pointing direction of the in-plane around $\mathbf{\Gamma}$ for the two cases, meaning it is associated with the distinct dielectric response we observe.

We will be focused on such opposite polarization since there may cause opposite 'spin torque' $\boldsymbol{\sigma} \times \boldsymbol{B}_R$ as we consider under applied electric field. Note that such radial polarized pseudo-spin can be attributed to the absent of both inversion and mirror symmetries and the presence of three-fold rotational axes of the crystal[35,36]. Similar pseudo-spin pattern we have observed in $CuInP_2S_6$ monolayer[37]. We can switch the interlayer stacking by inversion operation $P$ onto one layer with respect to the rotation center, when keeping the other layer fixed. For the H-type like and R-type like cases, their stacking potential switched with inversion symmetry operation. Consequently, there reverses in-plane components $\boldsymbol{B}_{eff,||}$ to $-\boldsymbol{B}_{eff,||}$, whereas out-of-plane component $\boldsymbol{B}_{eff,z}$ distribution rotates $30°$ with respect to $z$ axis.

### B. Generated spin current as response to Rashba SOC field

Upon applied perpendicular electric field, there induces the Rashba SOC term of pseudo-spin Hamiltonian $H_R = \alpha_R(k_x\sigma_y - k_y\sigma_x)$, where the coefficient $\alpha_R$ depends on the electric field $E_z$ for first-order perturbation. As $E$ evolves, $\alpha_R$ varies linearly as expected. Note that a vortex Rashba SOC field $\boldsymbol{B}_R$ (a pseudomagnetic field in momentum space) forms around $\mathbf{\Gamma}$. As a matter of fact, the electron states with inward and outward radial pseudospin polarization feel opposite 'spin torque' formulated as $\pm\boldsymbol{\sigma} \times \boldsymbol{B}_R$. And there gives the spin current correspondingly from below equation of motion:

$$i\hbar\frac{\partial\boldsymbol{\sigma}}{\partial t} = [\boldsymbol{\sigma}, H_R] \propto \boldsymbol{\sigma} \times \boldsymbol{B}_R \tag{1}$$

It is equivalent to a general continuity equation (current conservation) for spin component $\sigma^b$ in our case:

$$\frac{\partial \sigma^b}{\partial t} + \sum_{\alpha=x,y} \partial_\alpha J_\alpha^a = 0 \quad (2)$$

where the spin current tensor $J_\alpha^a \propto \alpha j_\alpha^a$ with $j_\alpha^a \propto \frac{\partial \sigma^b}{\partial t}$. According to electronic polarization $\boldsymbol{P}$ based on spin current mechanism[38], the general SOC can be formulated as $H_{SOC} = \lambda_{SOC} \sum_a J_\alpha^a A_\alpha^a$ with the gauge potential tensor $A_\alpha^a = \varepsilon^{a\alpha\beta} E_\beta$ and spin current tensor $J_\alpha^a = p_\alpha \sigma^a$. For the effective Lagrangian $L$ consists of $J_\alpha^a A_\alpha^a$, one has electric polarization $P_\beta \propto \frac{\partial L}{\partial E_\beta} \propto \varepsilon^{\beta a\alpha} \alpha j_\alpha^a$ and for our case:

$$P_x \propto y j_y^z, \quad P_y \propto -x j_x^z \quad (3)$$

where $j_x^z = j_y^z \propto (\boldsymbol{\sigma} \times \boldsymbol{B}_R)^z$. Since $j_\alpha^a$ depends on the planar wave vector $\boldsymbol{k}$ based on the pseudospin texture in the momentum space, the norm of above position vector $\boldsymbol{r} = (x, y)$ indicates the wavelength in the real space and its direction is along $\hat{\boldsymbol{k}}$. It is found that there forms vortex pattern for planar electric polarization. We can even obtain an in-plane electric susceptibility purely from SOC as $\partial P_\beta / \partial E_z \propto \pm \lambda_{SOC}$ due to Rashba SOC field $B_R \propto \alpha_R \propto \lambda_{SOC} E_z$, where '$\pm$' corresponds to R-type like and H-type like case respectively.

It should be noted that above discussion about pseudospin is based on only one general SOC Hamiltonian $H_{SOC}$. The spin current generated here we consider should be from the interaction between two different SOC sources i.e., the internal electric field and external electric field. However, the spin current mechanism that stems from this coupling effect is also based on $H_{SOC}$ as discussed. This implies that the source we consider for the coupling effect should be attributed to a self-consistent SOC field.

### C. SOC correction to the dielectric susceptibility

Without SOC, the planar dielectric polarization can be induced by applied electric field perpendicularly. It will be vanished as along as the mirror symmetry for *xy*-plane is present[34]. Thus, in the two stacking cases of $D_3$ symmetry, an in-plane dielectric response under out-of-plane electric field can be attributed purely to the geometry of the superlattice.

To be specific, for H-type like stacking, around the rotation center M, the nearest three S sites of the top layer turn clockwise twist with respect to Mo sites in the bottom layer for a $\theta$ angle, whereas for R-type like stacking, it turns out the three S and Mo sites twisted with the same direction but 60°-$\theta$ angle around center O. As for our example, with a large θ twist angle about 21.7°, the difference of the twist angle for the two stacking cases has tiny influence towards the dielectric polarization. And even though their twist center locates in different symmetric site, it tends to give nearly identical planar dielectric polarization without SOC. So far, we can understand the different dielectric response when SOC included comes from that within the system there redistribute charge to amplify or resist the external electric field. This mainly because, for the planar electronic polarization, the spin-current induced vortex distribution is opposite in the two distinct stackings. The amplified and suppressed dielectric response shown in the left panel of Fig. 1(b) stems from the opposite tendency to redistribute both intralayer and interlayer charge.

We can elucidate this effect and explain the DFT results from the SOC correction to the dielectric susceptibility. To be specific, we introduce a SOC dependent electrostatic term $-\lambda \hat{P}_{\beta} E_{\alpha}$ and treat it as perturbation when the external electric field $E_{\alpha}$ is along $\alpha$ direction (out-of-plane in our case):

$$H = H_0(E_{\alpha}) - \lambda \hat{P}_{\beta} E_{\alpha} \tag{4}$$

where the internal dipole in $\beta$ direction $P_{\beta}$ (in-plane in our case) can be coupled to the external electric filed $E_{\alpha}$ with a coefficient $\lambda$ via the spin current mechanism. $\lambda$ can be recognized as a renormalized coupling constant. If we consider the dielectric polarization $P_{\beta} = \chi_{\beta\alpha} E_{\alpha}$ with nonzero dielectric tensor $\chi_{\beta\alpha}$ comes from the structural symmetry of interlayer twist without SOC, then the renormalized tensor $\tilde{\chi}_{\beta\alpha} = (1+\lambda)\chi_{\beta\alpha}$ when SOC included. As we mention previously, this perturbation term should be equivalent to a self-consistent SOC field. In this way, $H_0(E_{\alpha})$ correspond to the Hamiltonian which only include non-self-consistent SOC term. Hence the perturbation term totally describes the self-consistent effect of SOC. These considerations can be carried in our DFT calculations.

As $E$ evolves from zero, we can expand the Bloch eigen states of $n$-th occupied band to $E_\alpha$ as $u_{n\boldsymbol{k},E_\alpha} \simeq u_{n\boldsymbol{k},0} + E_\alpha \partial_{E_\alpha} u_{n\boldsymbol{k},E_\alpha}|_{E_\alpha=0}$ . Then from the definition of electronic polarization based on BP method (in units of $e/(2\pi)^3$), for a band insulator[39]:

$$P_{n,\alpha}(E_\alpha) = \mathrm{Im}\int\left(\left\langle u_{n\boldsymbol{k},E_\alpha}\middle|\partial_{k_\alpha} u_{n\boldsymbol{k},E_\alpha}\right\rangle - \left\langle u_{n\boldsymbol{k},0}\middle|\partial_{k_\alpha} u_{n\boldsymbol{k},0}\right\rangle\right) dk_\alpha dk_\beta dk_\gamma \tag{5}$$

we can calculate that (to $E_\alpha$ order):

$$P_{n,\alpha}(E_\alpha) \simeq E_\alpha \int \Omega_{E_\alpha k_\alpha}|_{E_\alpha=0}\, dk_\alpha dk_\beta dk_\gamma \tag{6}$$

with the Berry curvature:

$$\Omega_{E_\alpha k_\alpha} = \mathrm{Im}\left(\left\langle \partial_{E_\alpha} u_{n\boldsymbol{k},E_\alpha}\middle|\partial_{k_\alpha} u_{n\boldsymbol{k},E_\alpha}\right\rangle - \left\langle \partial_{k_\alpha} u_{n\boldsymbol{k},E_\alpha}\middle|\partial_{E_\alpha} u_{n\boldsymbol{k},E_\alpha}\right\rangle\right) \tag{7}$$

We take $-\lambda \hat{P}_\beta E_\alpha$ term as perturbation and there gives the perturbated eigen states to first order of $E_\alpha$ as:

$$\left|u_{n\boldsymbol{k},E_\alpha}\right\rangle \simeq \left|u^{(0)}_{n\boldsymbol{k},E_\alpha}\right\rangle - \lambda E_\alpha \sum_{\varepsilon_{n',E_\alpha} > \varepsilon_F} \frac{\left\langle u^{(0)}_{n'\boldsymbol{k},E_\alpha}\middle|\hat{P}_\beta\middle|u^{(0)}_{n\boldsymbol{k},E_\alpha}\right\rangle}{\varepsilon_{n,E_\alpha} - \varepsilon_{n',E_\alpha}} \left|u^{(0)}_{n'\boldsymbol{k},E_\alpha}\right\rangle \tag{8}$$

where $\left|u^{(0)}_{n\boldsymbol{k},E_\alpha}\right\rangle$ and $\varepsilon_{n,E_\alpha}$ corresponds to eigen states and energy of $H_0(E_\alpha)$ , respectively. Then the dielectric susceptibility $\Omega_{E_\alpha k_\alpha}|_{E_\alpha=0}$ can be derived as:

$$\Omega_{E_\alpha k_\alpha}|_{E_\alpha=0} \simeq \Omega^{(0)}_{E_\alpha k_\alpha}|_{E_\alpha=0} + \tilde{\Omega}^{(0)}_{E_\alpha k_\alpha}|_{E_\alpha=0} \tag{9}$$

Here $\Omega^{(0)}_{E_\alpha k_\alpha}$ is the Berry curvature for zero-order states $u^{(0)}_{n\boldsymbol{k},E_\alpha}$ . And SOC correction can be derived as:

$$\tilde{\Omega}^{(0)}_{E_\alpha k_\alpha}|_{E_\alpha=0} = \frac{\lambda}{e}\mathrm{Re}\sum_{\varepsilon_{n',0} > \varepsilon_F} \frac{\left\langle u^{(0)}_{n\boldsymbol{k},0}\middle|\hat{P}_\alpha\middle|u^{(0)}_{n'\boldsymbol{k},0}\right\rangle\left\langle u^{(0)}_{n'\boldsymbol{k},0}\middle|\hat{P}_\beta\middle|u^{(0)}_{n\boldsymbol{k},0}\right\rangle + \left\langle u^{(0)}_{n\boldsymbol{k},0}\middle|\hat{P}_\beta\middle|u^{(0)}_{n'\boldsymbol{k},0}\right\rangle\left\langle u^{(0)}_{n'\boldsymbol{k},0}\middle|\hat{P}_\alpha\middle|u^{(0)}_{n\boldsymbol{k},0}\right\rangle}{\varepsilon_{n,0} - \varepsilon_{n',0}} \tag{10}$$

Consistent with DFT results, for R-type like stacking case, this correction $\tilde{\Omega}^{(0)}_{E_\alpha k_\alpha}|_{E_\alpha=0} > 0$ whereas for H-type like stacking one, $\tilde{\Omega}^{(0)}_{E_\alpha k_\alpha}|_{E_\alpha=0} < 0$ . We can infer that its sign is directly dependent on the sign of the coupling constant $\lambda$. And its magnitude

depends on $\langle \hat{P}_\alpha \rangle_{nn'} \langle \hat{P}_\beta \rangle_{n'n} + \langle \hat{P}_\beta \rangle_{nn'} \langle \hat{P}_\alpha \rangle_{n'n}$, which reflects a correlation between in-plane and out-of-plane electronic polarization in our case. The sum over occupied bands will give the total correction from SOC, which shows that the suppression of the dielectric susceptibility for H-type like case is so strong (right panel in Fig. 1(b)) from DFT calculations. As we consider, it mainly because of the much larger spin bands splitting in that case, which gives larger absolute value of coupling constant $\lambda$.

Finally, we should point out that the dielectric polarization associated with SOC effect in our case depends on the twist angle we choose. For larger twist angle approached to 30° shown in Fig. 3 there is almost no difference of the dielectric susceptibility for the two stacking cases. Under applied perpendicular electric field, it exhibits linear variation of the polarization $P$ with $E$, indicated by the calculated results for the two cases. On the contrary, for smaller twist angle about 13.2° (H-type like case) and 46.8° (R-type like case) we find a suppressed dielectric response of the former one compared with that of the latter, shown in Fig. 4. And the window for the $E$ field variation is $0 \sim 0.01$ eV/Å, where the polarization $P$ of H-type like case is even larger than that of R-type like case when $E = 0.01$ eV/Å. It is indicated that the dielectric suppression is weaker for this case of twist angle.

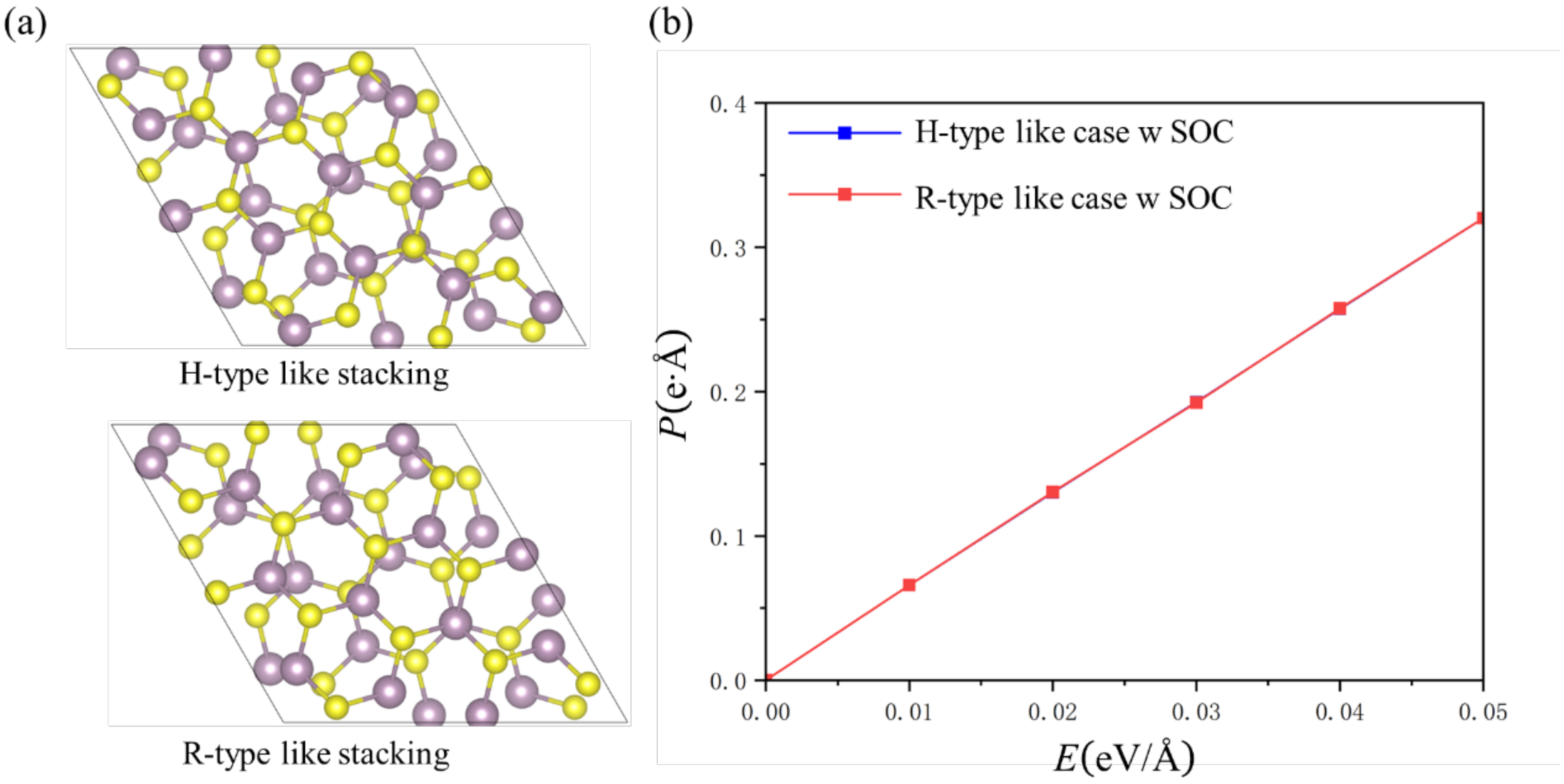

**Figure 3.** (a) Top view of 27.8° (H-type like) and 32.2° (R-type like) twisted bilayer $MoS_2$. The purple and yellow balls represent Mo and S atoms, respectively. (b) The BP method calculated electronic polarization $P$ variation as $E$ evolves from 0 to 0.05, in units of eV/Å.

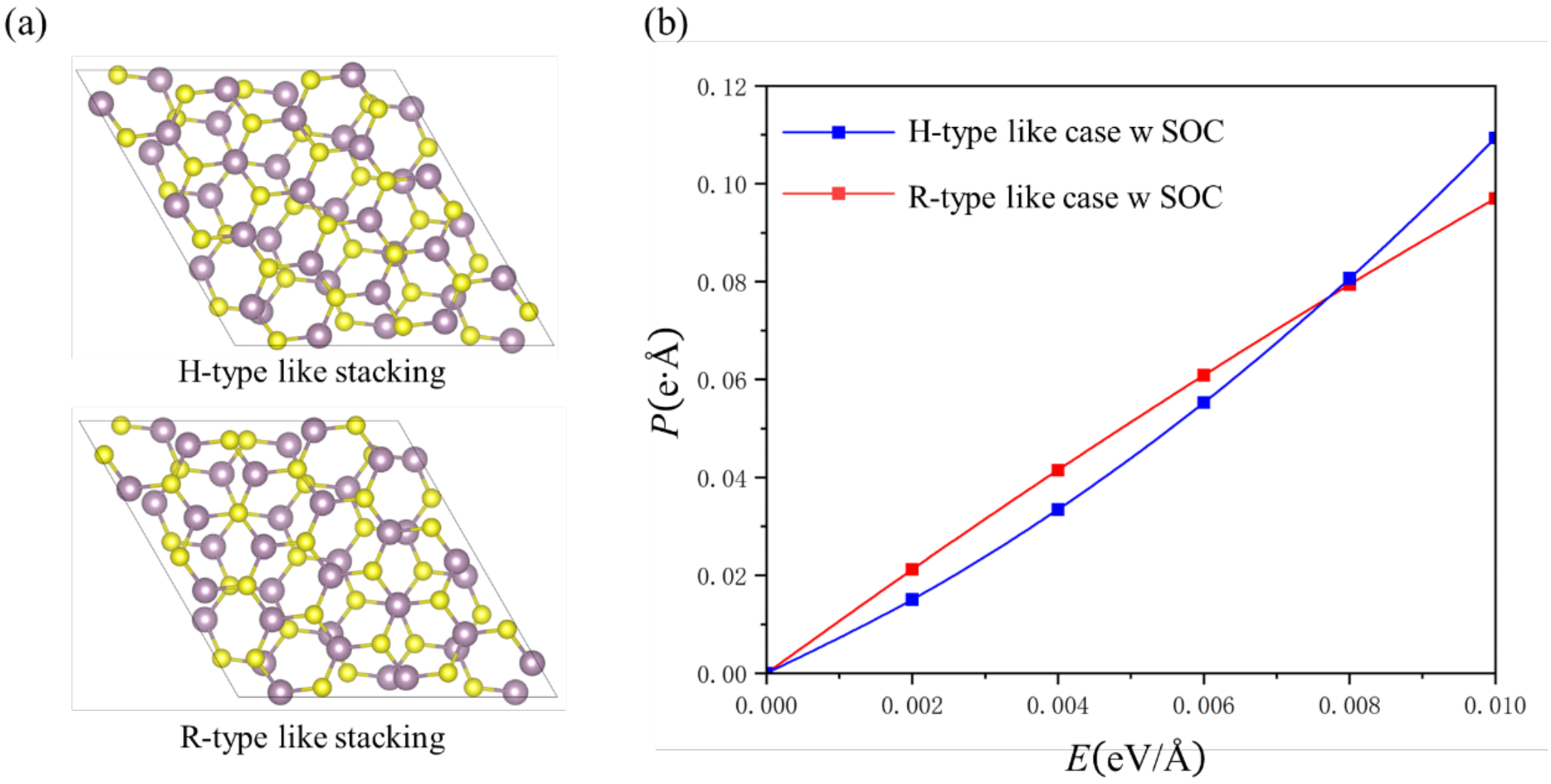

**Figure 4.** (a) Top view of 13.2° (H-type like) and 46.8° (R-type like) twisted bilayer $MoS_2$. The purple and yellow balls represent Mo and S atoms, respectively. (b) The BP method calculated electronic polarization $P$ variation as $E$ evolves from 0 to 0.01, in units of eV/Å.

To amplify the dielectric difference for smaller electric field applied, we suggest that the cooperation between the intrinsic SOC in the host bilayer and magnetism in the substrate may lead to considerable dielectric difference, similar to the mechanism in our previous work[17]. Then there even gives rise to total negative dielectric susceptibility. It is not our topic here and we leave it for future work.

## CONCLUSION

In summary, we show that the different stacking pattern of moiré system, defined as H-type like and R-type like case. Through calculating electronic polarization based on BP method in first-principles calculations with SOC included, it is demonstrated that under perpendicular electric field, there leads to positive and negative dielectric polarization contribution for R-type and H-type like stacking case, respectively. Our findings reveal a dielectric response in a magnetic way that through Rashba SOC, there modifies the internal pseudo-spin texture $\boldsymbol{\sigma}$ and generates an out-of-plane (pseudo-) spin current $\boldsymbol{j}_s \propto \boldsymbol{\sigma} \times \boldsymbol{B}_R$, subsequently resulting in opposite in-plane electric susceptibility for the two distinct stacking cases. Consequently, there forms opposite tendency to redistribute charge, ultimately leading to an amplified or suppressed dielectric response.

Moreover, the positive or negative in-plane electric susceptibility may suggest that it can be locked with a positive or negative 'magnetic charge' in 2D momentum space, respectively. Under an out-of-plane electric field, the magnetoelectric coupling gives rise to the formation of a vortex electronic polarization, indicating such a 'magnetic charge' possess dielectricity.

## ACKNOWLEDGMENTS

This work was supported by the National Key Research and Development Program of China (Grants No. 2022YFA1402902 and 2021YFA1200700), the National Natural Science Foundation of China (Grants No. 12134003 No. 12188101 No. 11834006), the excellent program in Nanjing University, and Innovation Program for Quantum Science and Technology (No. 2021ZD0301902). National funded postdoctoral researcher program of China (Grant No. GZC20230809), Shanghai Science and Technology Innovation Action Plan (No. 21JC1402000), ECNU Multifunctional Platform for Innovation.